\documentclass[floatfix,pre,superscriptaddress]{revtex4} 
\usepackage{latexsym}
\usepackage{graphicx}
\usepackage[tight,FIGTOPCAP]{subfigure}

\begin{document}

\date{\today}

\title{Randomization and Feedback Properties of Directed Graphs Inspired by
  Gene Networks.}

\author{M.~Cosentino Lagomarsino}
\affiliation{UMR 168 / Institut Curie, 26 rue d'Ulm 75005 Paris, France}
\affiliation{Universit\`a degli Studi di Milano, Dip.
    Fisica, Via Celoria 16, 20133 Milano, Italy } 
\email[ e-mail address: ]{mcl@curie.fr}
\author{B.~Bassetti} 
\affiliation{Universit\`a degli Studi di Milano, Dip.
    Fisica, Via Celoria 16, 20133 Milano, Italy } 
\affiliation{I.N.F.N., Milano, Italy} 
\email[e-mail address: ]{bassetti@mi.infn.it}
\author{P.~Jona} 
\affiliation{Politecnico di Milano, Dip. Fisica, Pza Leonardo Da Vinci
  32, 20133 Milano, Italy} 
\pacs{87.10+e,89.75.Fb,89.75.Hc}

\date{\today}

\begin{abstract}
  Having in mind the large-scale analysis of gene regulatory networks,
  we review a graph decimation algorithm, called ``leaf-removal'',
  which can be used to evaluate the feedback in a random graph
  ensemble.  In doing this, we consider the possibility of analyzing
  networks where the diagonal of the adjacency matrix is structured,
  that is, has a fixed number of nonzero entries.  We test these ideas
  on a network model with fixed degree, using both numerical and
  analytical calculations. Our results are the following.  First, the
  leaf-removal behavior for large system size enables to distinguish
  between different regimes of feedback. We show their relations and
  the connection with the onset of complexity in the graph.  Second,
  the influence of the diagonal structure on this behavior can be
  relevant.
\end{abstract}

\maketitle

\section{Introduction.}

Gene regulatory networks are graphs that represent interactions between genes
or proteins.
They are the simplest way to conceptualize the complex physico-chemical
mechanisms that transform genes into proteins and modulate their activity in
space and time.
In the network view, all these processes are projected in a static, purely
topological picture, which is sometimes simple enough to explore
quantitatively~\cite{UF05}.
%
Thanks to the systematic collection of many experimental results in
databases, and to new large scale experimental and computational
techniques that enable to sample these interactions, these graphs are
now accessible to a significant extent.  Some examples are the
undirected graphs of protein-protein interactions, and the directed
graphs of transcription and metabolic
networks~\cite{UF05,BLA+04,DRO+02,PRP04}.
%
The availability of such large-scale interaction data is extremely important
for post-genomic biology, and has provided for the first time a whole-system
overview on the global relationships among players in a living system.

The hope is to study these graphs together with the available information on
the genes and the physics/chemistry of their interactions to infer information
on the architecture and evolution of living organisms.
In this program, the simplest possible approach to take is to study
the topology of these networks. 
For instance, order parameters such as the connectivity and the clustering
coefficient have been considered~\cite{YOB04}.
Other investigators have focused on the relations of gene-regulatory graphs
with other observables, such as spatial distribution of genes, genome
evolution, and gene
expression~\cite{HPL04,Kep03,TB04,HCW04,LBY+04}

Typically, in an investigation concerning a topological feature of a
biological network, one generates so called ``randomized
counterparts'' of the original data set as a null model. That is,
random networks which conserve some topological observables of the
original.
The main biological question that underlies these studies asks to
establish when and to what extent the observed biological topology,
and thus loosely the living system, deviate from the ``typical case''
statistics.  To answer this question, the tools from the statistical
mechanics of complex systems are appropriate.
For example, a topological feature that has lead to relevant findings
is the occurrence of small subgraphs - or
``motifs''~\cite{MIK+04}.

The study presented here focuses on the topology, and in particular on the
problem of evaluating and characterizing the feedback present in the network.
On a generic biological standpoint, this is an important issue, as it is
related to the states and the dynamics that a network can exhibit.  Roughly
speaking, the existence of feedback in the network topology is a necessary
condition for the dynamics of the network to show multistability and
cycles~\cite{Tho73}.
In presence of feedback, the relations between internal variables play an
important role, as opposed to situations where the network is tree-like, and
the external conditions determine completely the configurations and the
dynamics.  Recently, we came to similar conclusions analyzing the structure of
the compatible gene expression patterns (fixed points) in a a Boolean model of
a transcription network~\cite{LJB05}. This model exhibits a transition between
a regime of simple gene control, and a regime of complex control, where the
internal variables become relevant and dynamically non-trivial solutions are
possible. These regimes correspond to the SAT, and HARD-SAT phases of
random-instance satisfiability problems.  For random Boolean functions, the
two regimes can be understood completely in terms of feedback in the network
topology. A selection of the Boolean functions can change this
outcome~\cite{CLP+06}.

Rather than dealing with specific experimental networks, this is meant
as a theoretical study on a model graph ensemble~\footnote{By the word
  ensemble, we mean here a family of graphs with a, typically uniform,
  probability distribution.}. Our purpose here is twofold.
First, to introduce some ``order parameters'', i.e.  functions that describe
the relevant feedback properties, connected to algorithms that can be used to
evaluate the feedback without enumerating the cycles.
Second, to study an ensemble of random graphs, or randomization
technique, with structured adjacency matrices, that conserve the
number of entries in their diagonal. This choice, which we will
justify, leads to a distinct behavior.
The two problems are introduced in section \ref{sec:prob}.  We show
the connections between different points of view on the problem, using
simple algebraic, graph theoretical, and statistical mechanical tools.
The first approach is an application of a decimation algorithm called
``leaf-removal''~\cite{BG01,MRZ03}.  This algorithm links the feedback
to the existence of a percolating ``core'' in the network, containing
cycles. The numbers of core variables and edges can then be used as order
parameters for the feedback. Here, we formulate three variants of the
leaf-removal algorithm, and discuss the statistical meaning and the
relations between them and different levels of feedback. Namely, for
an oriented graph, one can use these algorithms to define and
distinguish ``simple'' from ``complex'' feedback.
Furthermore, we discuss how one can connect feedback to the
satisfiability-like optimization problem of counting the solutions of
a random linear system on the Galois field GF2~\cite{Lev05}.
This can also be seen as a linear algebra problem concerning the kernel and
rank of the connectivity matrix. The theoretical motivation for the choice of
an ensemble with structured diagonal will follow naturally from this
discussion.
%
In section \ref{sec:res} we present our main results, as a series of
``phase diagrams'', which describe the typical feedback of random
realizations of the graphs.  In the unstructured case, the phase
diagrams obtained by leaf-removal show the existence of five regimes,
or ``phases'', characterizing the feedback in the limit of infinite
graph size. Some of these regimes are connected to complexity
transitions for the associated random GF2 optimization problem.
Moreover, we show that the choice of a structured diagonal leads to a
quantitatively different behavior, and thus to a significantly
different amount of feedback in the graph.
These differences are greatly enhanced if the degree distribution is
scale-free. 
%

\section{Formulation of the Problem and Algorithms}
\label{sec:prob}

\begin{figure}[htbp]
  \centering
  \includegraphics[width=0.25\textwidth]{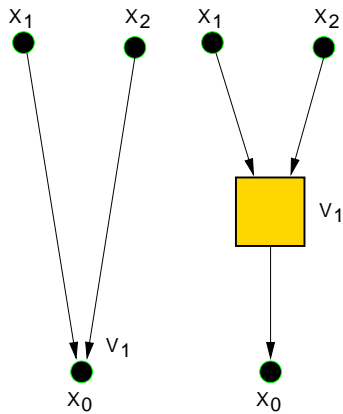}
  \caption{Different representations of interactions in
    the graph $G$. Left: oriented graph Right: a bipartite oriented graph.
    $V_{1}$ is a function and $x_{i}$ are variables, $x_0$ is the output.}
  \label{fig:nodes}
\end{figure}

The problem we want to address consists in evaluating the feedback in a random
ensemble of graphs. While the range of application is more general, to avoid
excess of ambiguity we choose a specific ensemble of graphs that will be
treated in detail throughout the paper.
We consider oriented graphs, where each node has $p$ incoming links.
The graph ensemble can be specified through a $M \times N$ Boolean
matrix $B$ (having elements $0$ or $1$). $B$ represents the
input-output relationships in the network. If $x_{i}$ are network
nodes, $B_{ji} = 1 $ if $x_{i} \rightarrow x_j$, and zero otherwise.
The matrix is rectangular because only $M<N$ nodes have an input.  We
allow for self links, or diagonal elements.
For a simple directed graph one can say that feedback exists as soon
as closed paths of directed edges emerge.  
%
Having in mind the fact that, while here we consider only topological
properties, the incoming links are ``inputs'', that is, they encode for
some conditions on the nodes (for example, on gene expression), we can
also use a separate graphical representation for the nodes, or
``variables'', and the ``functions'' regulating these variables. This
representation is a bipartite graph (Fig.~\ref{fig:nodes}). Each graph
has $N$ variables and $M$ functions, and thus on average $\gamma= M/N$
functions per variable.

An important point concerning randomization, is that the choice of
what feature to conserve and what not to conserve in the randomized
counterpart is quite delicate and depends on specific considerations
on the system.  In the words of statistical mechanics, the typical
case scenario can vary greatly with the choice of the ensemble.  For
instance, the network motifs shown by randomizing a network with an
Erdos-Renyi random graph differ from the usual ones, for which the
degree sequence is used as a topological invariant~\cite{IMK+03}.
In studies of biological networks, the diagonal of $B$ is normally
disregarded, or assumed to have the same probability distribution as a
row or a column.  The use of considering it is a matter of the nature
of the graph and the property under exam.
For the case of transcription networks, an ensemble with structured diagonal
might have some relevance.  For example, for motifs discovery, sometimes one
puts the diagonal to zero, and considers degree-conserving randomizations that
do not involve the diagonal~\cite{SMM+02}.
%
In our earlier work on transcription networks, we have considered the
autoregulators as a structured diagonal~\cite{LJB05}. We will show,
for our model graph ensemble, that this leads to considerably
different results for the feedback.  There are other biological
examples where a structured adjacency matrix emerges naturally. The
simplest example are mixed interaction graphs. For instance, one can
consider a composition of a transcription network with a
protein-interaction network (which is a non directed graph) and pose
the question of evaluating the feedback on a global scale compared to
randomized counterparts.

\paragraph{Leaf-removal algorithms.}
\label{sec:lr}

\begin{figure}[htb]
  \centering
\includegraphics[width=\textwidth]{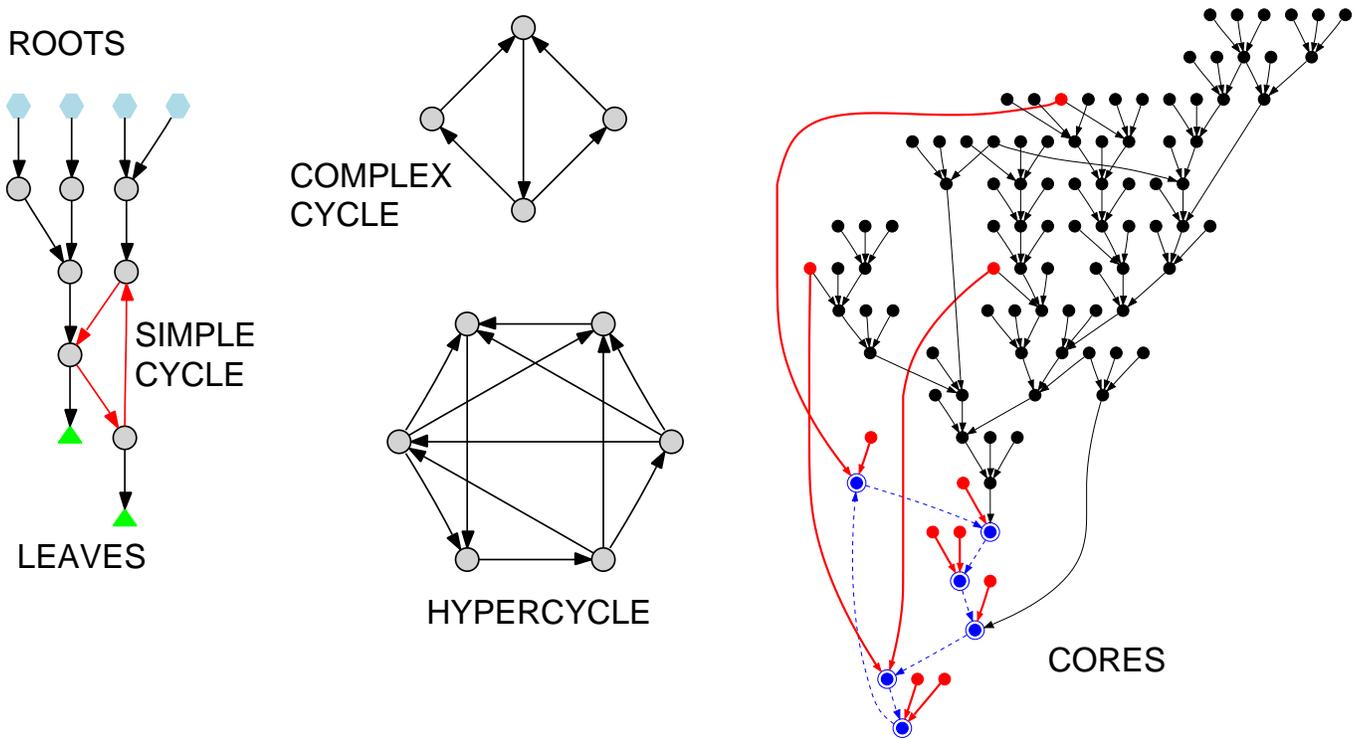}
\caption{Left: example of roots (free variables) and leaves for the
  leaf-removal algorithm. This graph contains a simple cycle (in red), which
  is not removed by LRa and LRb, but is removed by LRc. Middle: examples of a
  complex cycle and a hypercycle. A complex cycle (top) is not removed by LRc,
  but does not belong to the kernel of of $A^{t}$.  A hypercycle (bottom) is
  an element of the kernel of $A^{t}$, because each variable appears in an
  even number of functions.  Right: example of cores for the different
  leaf-removal variants, applied on the same initial graph.  The image refers
  to a random graph with $p=3$, $\gamma = 0.5$, $N=600$. The cores are
  represented as a directed graph, and superimposed.  The LRa core (whole
  figure) contains feedback loops and tree-like regions (black) upstream of
  the loops.  The LRb core (red) does not contain the treelike parts, but all
  the feedback is preserved.  The LRc core is empty, as this algorithm is able
  to break simple cycles connected to single free variables.  The cycle of the
  original graph is indicated by circled nodes and dashed edges (blue).}
  \label{fig:cores}
\end{figure}

A straightforward way to measure the amount of feedback in a graph is
to count cycles.  However, this is in general computationally as
costly as enumerating all the paths. For this reason, it is desirable
to use algorithms and order parameters that allow a quicker
evaluation.
To this aim, we describe three variants of a decimation algorithm,
termed ``leaf-removal'', that is able to remove the tree-like parts of
the graph, leaving the components with feedback.
We define a leaf as a variable having only incoming links, and a
``free'' variable, or a root, a variable having only outgoing links
(Fig~\ref{fig:cores}).  $\gamma$ is a measure for the fraction of
regulated variables, as opposed to external variables which only enter
functions as inputs.  The three variants of the leaf-removal
iteratively remove links and nodes from the graph, using the following
prescriptions (Fig~\ref{fig:cores}).
\begin{enumerate}
\item LRa. Remove leaves and their incoming links.
\item LRb. As above. Additionally, remove incoming links of nodes whose
  incoming links are all connected to roots, which are also removed.
\item LRc. As LRa. Additionally, remove all the incoming links
  (together with their associated nodes) of nodes
  whose incoming links are connected to at least one root.
\end{enumerate}
This is an iterative nonlinear procedure, where more variables may
disappear in a single move. 
LRc works naturally on directed and undirected bipartite graphs.  In
fact, viewing the system as a bipartite graph, one can verify that LRc
is equivalent to removing all the functions connected to a single
node, ignoring directionality.
Instead, LRa and LRb are thought for a directed graph, such as the ones
we consider here. 

There are two possible outcomes for the leaf-removal. Removing the
whole graph, or stopping at a core subgraph that contains cycles.
The core is composed of $N_C$ genes and $M_C$ functions. We want to use these
as order parameters for the feedback.  Equivalently, we can use $\Delta_C =
\frac{N_{C}- M_{C}}{N}$
and $\gamma_{C} = M_{C}/N_{C}$.
The difference between LRa and LRb is that LRb is able to remove
tree-like parts of the graph that are upstream of a simple cycle. LRc
is also able to do this.  On the other hand, LRc might break some of
these cycles because it disregards the orientations of the edges
(Fig.~\ref{fig:cores}). LRc cannot break ``complex'' cycles, defined as cycles
where each node is connected to at least two functions. 


\paragraph{Connections with Random Systems in GF2 and Adjacency Matrix Algebra.}
To investigate the feedback properties of the graph, one can also consider the
following linear system in the Galois field $\mathrm{GF2}$ (the set $\{0,1\}$
with the conventional operations of product, and sum modulo 2).
\begin{equation}
\label{eq:syst}
 Ax = v \ \ .
\end{equation}
Here, $v$ is a random vector of $\mathrm{GF2}^M$, that represents the
functions, and $A = B + I_{MN}$, where $I_{MN}$ is the truncated $M \times N$
identity matrix, and the sums are in $\mathrm{GF2}$.  In other words, we
imagine that each output variable is subject to a random XOR constraint, and
the idea is to use this as a probe for the feedback. Each XOR constraint, or
GF2 equation corresponds to a function. In the language of statistical
mechanics, the random linear system (\ref{eq:syst}) maps to a p-spin model on
the graph~\cite{MRZ03}.
The important point is that feedback translates into algebraic properties of
the matrix $A$ in GF2, and in solutions of Eq.~(\ref{eq:syst}).  A feedback
loop, or a cycle, corresponds to the pair $A^{o},h^{o}$, where $A^{o}$ is a $l
\times l$ submatrix of $A$, and $h^o$ is a $l$-component vector such as $h^o
A^{o} = 0$. Indeed, the functions and variables selected by the nonzero
elements of $h^o$ are such that each variable appears in an even number of
constraints.

We can also define a ``hypercycle'' as an $M$ component vector $h$ of
GF2, such as the right product $hA = 0$, because the functions and
variables selected by the ones in $h$ are such that each variable
appears in an even number of functions. Graphically, a hypercycle is a
connected cluster made of functions that share an even number of nodes
(Fig.~\ref{fig:cores}). From the algebraic point of view, it is an
element of the kernel of $A^{t}$, and is then connected to the
solvability of Eq.~(\ref{eq:syst}).
This consideration enables to evaluate the average number
$\overline{\cal N}$ of solutions of Eq.~\ref{eq:syst}. Perhaps
surprisingly, one can prove that $\overline{\cal N} = 2^{N-M}$ under
very general conditions.  However, this average ceases to be
significant when the hypercycles become extensive (i.e., the number of
nodes they involve has order $N$), as the fluctuations become
dominant.  This is discussed in detail in
Appendix~\ref{sec:randomgf2}.
The exact threshold for $\gamma$ where hypercycles become extensive is a phase
transition in the thermodynamic limit $N \to \infty, \ \ M \to \infty$ at
constant $\gamma$. Precisely, it is called the SAT-UNSAT transition for
Eq.~(\ref{eq:syst})~\cite{MPZ02}.
The UNSAT threshold depends on the graph ensemble, and has been
determined in some cases~\cite{Kol98}. In some instances, there may
exist also an intermediate ``HARD-SAT'' or glassy phase, where
$2^{N-M}$ solutions exists, but they belong to basins of attractions
whose distance from each other~\cite{MPZ02} is order $N$. For a p-spin
problem on a graph, this glassy phase corresponds to the presence of
complex cycles~\cite{MRZ03}.

\paragraph{Structured diagonal.}
As a hypercycle is a particular realization of a complex cycle, it is easy to
understand how the core of a leaf-removal algorithm will in general (but not
always) contain hypercycles: none of the algorithms is able to break these
structures.  
%
This is shown in Appendix~\ref{sec:matrA}, which discusses the relation of the
leaf-removal ``moves'' with operations on the rows and columns of $A$, related
to the solution of Eq.~(\ref{eq:syst}).  As explained there, for a directed
graph, the extensive hypercycle, or UNSAT region may exist only at $\gamma=1$.
In the case where the diagonal is structured, the situation is quite
different, and the hypercycle phase can appear for
$\gamma<1$~\cite{LJB05,CLP+06}.
The above consideration justifies from an abstract standpoint the intermediate
situations, with a fixed fraction of ones on the diagonal of $A$.  In
considering this ensemble of matrices with structured diagonal, we can
introduce an additional parameter $\chi$, that represents the fraction of ones
on the diagonal of $A$.
It is important to note that the introduction of a structured diagonal in $A$
changes the adjacency matrix, and thus the graph ensemble.
This change can have different interpretations. Rather than focusing on a
particular one, the objective here is to show on an abstract standpoint how
the phase behavior of Eq.~(\ref{eq:syst}) is perturbed by $\chi$.

\section{Regimes of Feedback}
\label{sec:res}


In this section, we discuss numerical and analytical results for the
leaf-removal algorithms that support the general considerations above.
We considered mainly the ensemble of graphs with fixed indegree $p$ and
Poisson-distributed outdegree $k$, $p(k) = \frac{(p \gamma)^{k}}{k!}
e^{-p\gamma}$. The diagonals are thrown with independent probability, to
ensure that the average fraction of ones is $\chi \in [0,1]$.  The choice of a
structured diagonal does not perturb the marginal probability distributions of
columns or rows. 
One can connect $\chi$ to the notion of ``orientability''.  If $M>N$, it is
impossible to orient a graph assigning one single output per function. On the
other hand, a graph with a structured diagonal can be seen as a partially
oriented one, where some directed constraints coexist with some undirected
ones.  In this interpretation $\chi=1$ is the simple directed graph with no
self-links. The case $\chi=0$ can be seen as a totally undirected graph, a
similar ensemble to that used in~\cite{MRZ03}.

\begin{figure}[htb]
  \centering
  \includegraphics[width=0.6\textwidth]{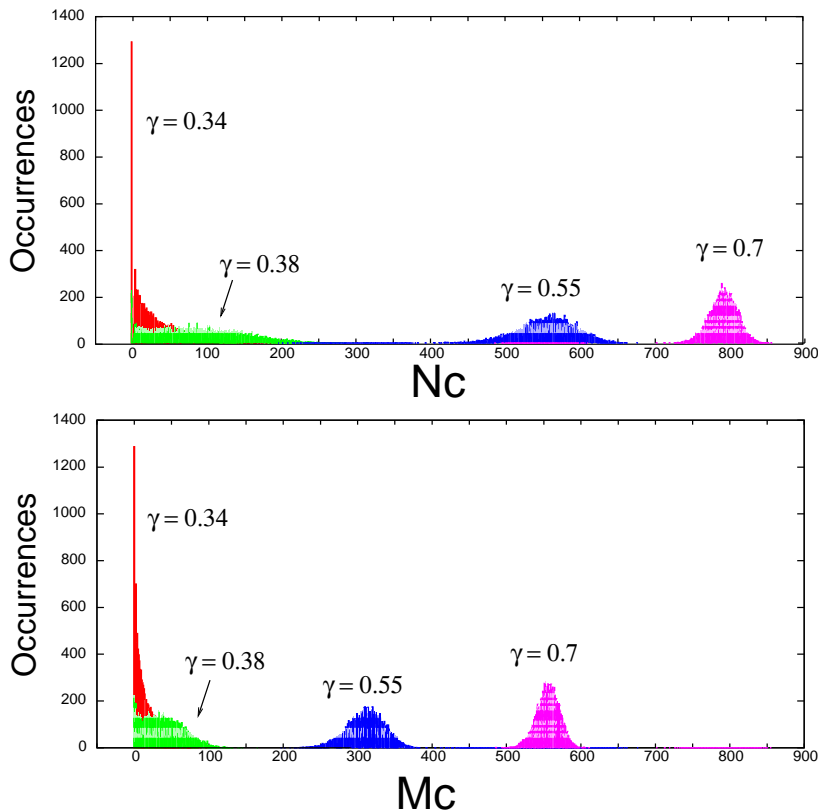}
  \caption{Histogram of the core dimensions $N_{C}$ and $M_{C}$ as a function
    of $\gamma$ for LRb. The data refer to $10^4$ random networks with $p = 3$
    and initial size $N = 1000$.  For low $\gamma$, the cores are clustered
    towards the empty graph.  At $\gamma \simeq 0.38$ the core distribution
    becomes wide.  Successively, the mean values grow and the histogram
    acquires again a sharp single peak at increasing $M_{C}, N_{C}$. This is
    reminiscent of a second order phase transition. For LRc, this transition
    is much sharper (first order), and marks the onset of complexity in the
    core solutions $\gamma_{d}^{c}$.}
  \label{fig:histo}
\end{figure}

We start with the totally orientable case $\chi = 1$.  For each value of
$\gamma$, at fixed network size $N$, one can generate randomized graphs and
evaluate their cores numerically. This procedure is exemplified in
Fig.~\ref{fig:histo} for the case of LRb.  The figure shows a transition to a
regime where the core is nonempty and all the graphs are sharply distributed
around the average core size.  Equivalently, one can evaluate the core order
parameter $\Delta_C$, which vanishes when the core is empty or $M_{C} =
N_{C}$.  The same order parameter is negative when $M_{C} > N_{C}$.
Each LR has two critical values. The first, $\gamma_{d}^{x}$, is
associated to the emergence of a nonempty (extensive) core. The second
$\gamma_{s}^{x}$, to the condition $N_{C}<M_{C}$. Based on our
results, $\gamma_{s}$ is always the same for all three leaf-removals,
and corresponds to the onset of the UNSAT phase of extensive
hypercycles. From simulations and analytical work, $\gamma_{s} = 1$.
$\gamma_{d}^{x}$, instead, depends on the ability to remove parts of
the graph of the different algorithms.

\begin{figure}[htbp]
  \centering
\includegraphics[width=\textwidth]{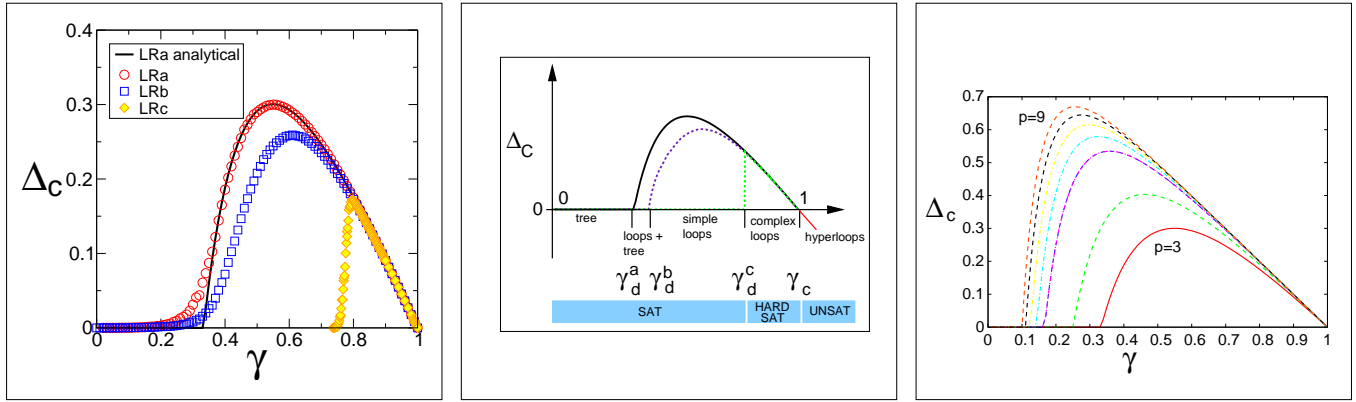}
  \caption{Left: $\Delta_C(\gamma)$ for $\chi=1$, $p=3$. The solid line
    corresponds to the analytical calculation (Appendix~\ref{sec:LRa}). The
    symbols are numerical results for $10^{3}$ realizations of graphs with
    $N=1000$. $\gamma_{d}^{a} < \gamma_{d}^{b} < \gamma_{d}^{c}$ mark the
    transition to an extensive core for the three leaf-removal algorithms.
    $\gamma_{s} = 1$ for all three algorithms to the point where $\Delta_C$
    becomes negative. Middle: A scheme of the resulting phase diagram. Right:
    Analytical ($N \to \infty$) values of the order parameter $\Delta_{C}$ for
    the LRa algorithm, $\chi=1$ and different values of $p$.  The order
    parameter deviates from zero at the threshold $\gamma_{d}^{a} = 1/p$, and
    crosses again at $\gamma_{c}=1$. The calculation is described in
    Appendix~\ref{sec:LRa}.  }
  \label{fig:deltavari}
\end{figure}

As we have seen, LRa can remove less than LRb, because the latter is able
to deal with the tree-like parts of the graph that lay upstream of the loops.
Also, LRb can remove less than LRc, because LRc can break feedback loops if
they are connected to a single free variable.  Thus, one can expect
$\gamma_{d}^{a} < \gamma_{d}^{b} < \gamma_{d}^{c}$.  This is indeed our
observation (Fig.~\ref{fig:deltavari}).

Based on these results, we can distinguish the following five regimes of
feedback: (1) all cores are empty, (2) only the LRa core is nonempty, (3) both
the LRa and the LRb core are nonempty, (4) all the cores are nonempty with
$N_{C}>M_{C}$, (5) all the cores have $N_{C}<M_{C}$.  These last two regimes
can be seen as thermodynamic phases connected with the SAT-UNSAT transition of
the associated linear system.
\begin{enumerate}
\item There are no feedback loops in the typical case.
\item Feedback loops emerge, that form a core having an extensive
  treelike component upstream. The cycles are intensive (i.e. the core
  contains a number of nodes negligible with respect to N, or $o(N)$),
  but the tree upstream becomes extensive ($O(N)$).  Analytically, one can
  compute that $\gamma_{d}^{a}$ corresponds to the percolation-like
  threshold $1/p$ (see Appendix \ref{sec:LRa} and
  Fig.~\ref{fig:deltavari}). Intuitively, as soon as the graph
  percolates, even in the presence of a small region containing
  cycles, the tree upstream of the feedback loops can span an
  extensive part of the graph.
\item There is an extensive core of simple loops. LRb erases the tree upstream
  of the feedback loops, thus it can only have its threshold when the region
  of cycles itself becomes extensive.  So far, we have not been able to
  compute the threshold $\gamma_{d}^{b}$ analytically.  However, our
  simulations indicate that it lies higher than $\gamma_{d}^{a}$
  (Fig.~\ref{fig:deltavari}).
\item HARD-SAT phase. Intensive hypercycles, and extensive complex
  cycles form the core, where each variable appears in 2 or more
  functions. This gives a clustered structure to the space of
  solutions in the corresponding random linear system. $\Delta_C$ is
  proportional to the complexity $\Sigma$ of the space of solutions,
  defined by the relation $ \log\mathcal{N} \sim N (\Sigma + S)$. Here
  $S$, the entropy, measures the width of each cluster, while $\Sigma$
  counts the number of clusters.
\item UNSAT phase. The hypercycles become extensive. The threshold $\gamma_{s}
  = 1$ can be compute analytically (see Appendix~\ref{sec:LRa}, and
  Fig.~\ref{fig:deltavari})
\end{enumerate}

%
\begin{figure}[htbp]
  \centering
  \includegraphics[width=\textwidth]{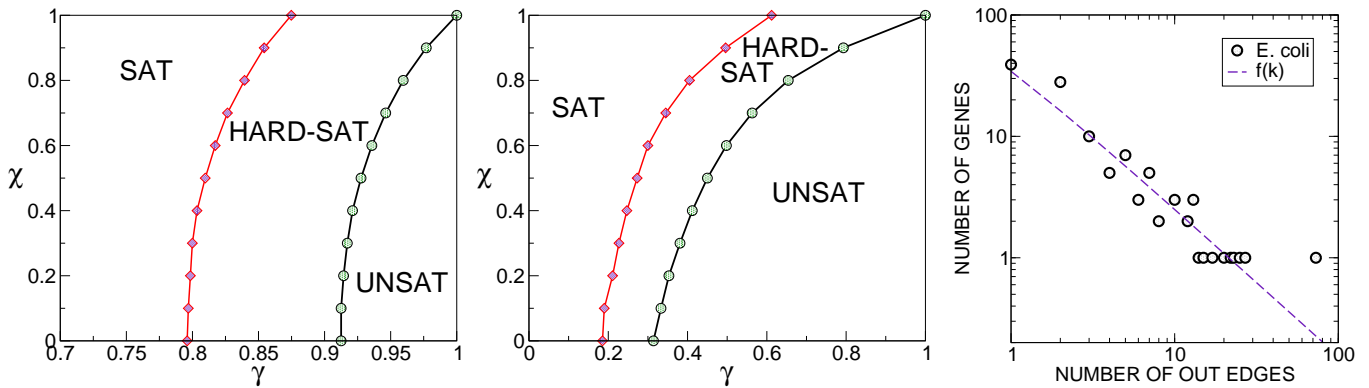}
  \caption{Left: Phase diagram for $p=3$ and structured diagonals (varying
    $\chi$). There are quantitative changes with respect to $\chi = 1$.
    Middle: Phase diagram for scale-free distribution of the outdegree $k$.
    $\gamma_{d}^{c}$ and $\gamma_{s}$ move with the same trend and undergo a
    notable quantitative drift with increasing $\chi$. Right: The exponent for
    the outdegree distribution is a fit from data on the transcription network
    of E.~coli~\cite{SMM+02}.}
  \label{fig:phasedplaw}
\end{figure}

Considering now ensembles with a structured diagonal, one can carry the same
analysis at fixed values of $\gamma$ and $\chi$.  As we discussed above, LRc
is not sensitive to graph orientation, and graphs with a structured diagonal
can be seen as partially oriented ones. Thus, the simplest choice is to forget
the other variants of the algorithms and focus on LRc.  At fixed $\chi$, there
are three phases SAT, HARD-SAT, and UNSAT. On the other hand, as we argued
above, because of the structure of the core matrices, these regions vary with
$\chi$, and a new phase diagram can be generated.  The interesting result is
that this ensemble can show quantitatively different thresholds, while leaving
the marginal distributions for the row and column connectivity unchanged.  We
have addressed this question numerically, computing the thresholds
$\gamma_{d}^{c}(\chi)$ and $\gamma_{s}(\chi)$. The results for the fixed $p$
ensemble are shown in Fig.~\ref{fig:phasedplaw}. The value for both
thresholds increases with increasing $\chi$. In particular, $\gamma_{s}(\chi)$
becomes exactly $1$ in the directed case. On the other hand, the phenomenology
of the transition does not vary with $\chi$, with a discontinuous jump at the
onset of a complex cycles phase, as in a first order phase transition.
Thus, in the fixed $p$ ensemble, there is a marked quantitative change in the
thresholds.  One may wonder whether the impact is the same for ensembles of
graphs where the connectivity distributions are wider.  Throughout the paper
we have considered only the ensemble with fixed $p$ and Poisson distributed
$k$.  Notably, the effect of a structure diagonal becomes larger for
scale-free distributions of $k$. This is illustrated in
Fig.~\ref{fig:phasedplaw}, where we show the phase diagram for a power-law
distribution for $k$ with exponent $1.22$ fitted from data from
E.~coli~\cite{SMM+02}, and independently thrown columns for $A$.  In this
case, the influence of the diagonal can bring the hypercycle threshold
$\gamma_c$ down by a factor of three.

%

\section{Discussion and Conclusions}
\label{sec:cooncl}

We presented a theoretical study focused on the evaluation of feedback
and the typical behavior of graphs taken from a random ensemble.  The
study focuses specifically on the ensemble of directed graphs with
fixed indegree and Poisson outdegree. On the other hand, it is
inspired by examples of biological graphs.
Detecting feedback in large biological graphs and their randomized
counterparts is important to understand their functioning. The use of our
technique is that it allows for a quick evaluation and, more importantly, it
provides some quantitative large-scale observables that can be used to measure
the weight and the complexity of feedback loops.
In order to do this, we introduce different variants of the leaf-removal
algorithm, which naturally carry the definition of simple order parameters,
depending on the properties of the core.
We showed how the three algorithms relate to graph properties, algebraic
operations on the adjacency matrix, and to solutions of the associated linear
systems of equations in GF2.  This analysis naturally leads to the abstract
introduction of structured random graphs that conserve the number of entries
in the diagonal of the adjacency matrix, which might be relevant in some
biological situation. 

Our two main results are the following.  First, a phase diagram of different
regimes of feedback depending on the fraction of free variables for an
oriented graph. It shows a quite rich behavior of phase transitions that are
interesting from the statistical physics viewpoint.  These include the
thresholds observed in diluted spin systems and XOR-like satisfiability
problems. As already observed in~\cite{MRZ03}, the onset of the complex phase
is deep in the region where cycles exist and they involve a subgraph of the
order of the graph size.
On the other hand, the less intricate feedback regimes of intensive simple
cycles connected to extensive trees, and of extensive simple cycles, might be
relevant to characterize the dynamics in biological instances.  The
leaf-removal algorithms enable to analyze these different forms of feedback,
that can be ``weaker'' than the complex cycles and hypercycles that are
relevant for the associated GF2 problem.
The second result is that the introduction of a structured diagonal,
which can be interpreted as a partial orientation in the graph, has
some influence on the thresholds.  This is particularly true in
presence of scale-free degree distribution, where we showed a phase
diagram inspired by the connectivity in the E.~coli transcription
network~\cite{SMM+02}.
The algorithms described here can be readily applied to biological data sets
and their randomized counterparts. We are currently addressing this question
in relation with the Darwinian evolution of some transcription and mixed
transcription- and protein-interaction graphs.  Finally, while this analysis
is loosely inspired to graphs related to gene regulation, the need to evaluate
the feedback arises in different contexts, where the tools described here
could prove useful.



\begin{appendix}

\section{Appendix}

\subsection{Solutions of the Random System in GF2}
\label{sec:randomgf2}

Evaluating the average number of solutions of Eq.~\ref{eq:syst} for large $N$
at constant $\gamma$ gives information in the feedback of the associated
graph. We denote the kernel of a matrix by $K$, its range by $R$, and their
dimensions by $\kappa$ and $\rho$ respectively.  If the probability measure
for $v$ is flat, the average number of solutions for fixed $A$ is the
probability that $v \in R(A)$, i.e.
\begin{displaymath}
\mathrm{prob}(v \in R(A)) = \frac{2^{\rho(A)}}{2^M} = 2^{- \kappa(A^t)} \ , 
\end{displaymath}
times the number of elements in $K(A)$ (i.e. $2^{K(A)}$.  The average number
of solutions is thus
\begin{equation}
  \label{eq:sols}
  \overline{\cal N} = \langle  2^{- \kappa(A^t)}  2^{\kappa(A)}\rangle_{A}  =
  2^{N-M} \ \ ,
\end{equation}
where we have used the relations $\rho(A) + \kappa(A) = N$, $\rho(A^t) +
\kappa(A^t) = M$, and $\rho(A) = \rho(A^t)$.
Moreover, with the same reasoning, the fluctuations in the number of
solutions are
\begin{displaymath}
  \overline{{\cal N}^{2}} =(\overline{\cal N})^2 \langle 2^{\kappa(A^t)}
  \rangle_{A} \ ,
\end{displaymath}
meaning that when the average $ \langle 2^{\kappa(A^t)} \rangle_{A}$ is
$O(1)$, an average number of solutions $\overline{\cal N} = 2^{N-M}$ are
typically found, while this is not the case if $ \langle 2^{\kappa(A^t)}
\rangle_{A}$ is an extensive quantity. In fact, when this ``selfaveraging''
property breaks down, typically no solutions are found, because
$\overline{\cal N}$ is supported only by the multiplicity of very rare $ v \in
R(A)$.  This connects the solvability of the system to the topology of the
hypercycles.

There are phase transitions between the two above regimes, tuned by the order
parameter $\gamma$. The standard approach is to take the thermodynamic limit
$N \to \infty, \ \ M \to \infty$ at constant $\gamma$. These transitions
depend on the ensemble of graphs considered~\cite{Kol98}.

\subsection{Adjacency Matrix and Leaf-removal.}
\label{sec:matrA}



Let us try to visualize the leaf-removal procedure, for instance LRc, on a
generic adjacency matrix.  Consider a general Boolean matrix $A$ $M\times N$,
and apply LRc. Each time we find a leaf, we assign it and its corresponding
constraint a progressive number, and we use that number as a label for the
rows.  With these permutations, we construct a hierarchy for the leaves, as
the leaves of layer $a$ cannot appear in the clauses of layer $b \ge a$.  In
the tree-like case, reordering the lines of $A$, we obtain
\begin{displaymath}
\left( \begin{array}{c||c|c|c|c|c||c|c}
\mathrm{layer}& N& ...&...&...&...& N-M&1\\
\hline
(1) &I &...&...&...&...&...&...\\
(2) &0&I&...&...&...&...&...\\
(...) &0 &0&...&...&...&...&...\\
(m-1) &0&..&0&I&...&...&...\\
(m) &0 &0&..&0&I&...&...\\
\end{array} \right)
\end{displaymath}
where $(1)$ is the set of first layer leaves, $(2)$ the second, etc.  The last
$N-M$ entries of each row correspond to free variables. We have thus obtained
a triangulation of $A$, where the diagonal is made of blocks (the layers) of
identity matrices.

In the presence of a core, the triangulation can be carried only until a the
core is reached, and the the matrix can be rearranged to show the core in the
lower right corner.
If the core has hypercycles, in the UNSAT phase, the matrix structure is
\begin{displaymath}
\left( \begin{array}{c||c|c|c|c||c} 
\mathrm{layer}&N ...&...& ... & ... & N_c \leftarrow \rightarrow 1\\
\hline
(M) &1! &...&....&...&...\\
(..) &0&1!&...&...&...\\ 
(...) &0 &0&1!&...&...\\
\hline 
(M_c)&0&..&0& 0!& \mathrm{core}     \\ 
(..) &0&..&0& 0 & `` \\ 
(1)  &0&..&0& 0 & `` \\ 
\end{array} \right) 
\end{displaymath}
Here, $M_C > N_C$, so typically it will not possible to find solutions to the
core linear system on GF2, or the core does not contain sufficient free
variables. 
When the ensemble for $A$ is specified, one has to apply this procedure to all
the realizations.
Naturally, the outcome depends on the matrix ensemble. It also depends in
general on the variant of leaf-removal that one applies.

\paragraph{Structured diagonal.}
\label{sec:strdiag}
Focusing on the diagonal of $A$, we note that in presence of hypercycles, one
has necessarily to have some zeros in the $M \times M$ submatrix of $A$ to
realize the condition $N_c<M_c$. This can be seen in the sketch above, where
the diagonal elements are followed by an exclamation mark.
In particular, the diagonal contains an extensive number of ones. Thus,
following the above argument, it is easy to realize that $M_c \le N_c$, and
the hypercycle phase may exists only marginally at $\gamma=1$.  For our main
choice of ensemble, this is the case, as each variable can have only one
input, so each constraint can always be labeled by the name of its output
variable, which will appear as a one in the diagonal of $A$.
In the case where the diagonal contains an extensive number of zeros,
the situation is quite different, and the hypercycle phase can appear
for $\gamma<1$~\cite{LJB05,CLP+06}.

\subsection{Analytical results for LRa}
\label{sec:LRa}


We present here the analytical calculation for LRa.  If $f_k$ is the
probability to have $k$ outputs, LRa defines a dynamics for it,
associated by the cancellations of leaves at each time step. For every
time $t$, one can write
$$
\begin{array}{l}
 N = N \: \sum f_k(t) \ ; \vspace{0.1cm} \\
 N(t)= N\: \sum_{k\geq 1} f_k(t)= N\: (1-f_0(t)) \ ; \\
\vspace{0.1cm}
 M(t)\:p = N \: \sum k f_k(t) \ .
\end{array}
$$
The fraction of nonempty columns is given by the probability $1-f_0(t)$.
Writing the increments as, $\Delta N_k =N \:\Delta f_k = N\: \frac{\partial
  f_k}{\partial t} \Delta t$, one can choose $\Delta t= \frac{1}{M}$, $t \in
[0: 1]$, and obtain intensive equations of the kind
$
\frac{\partial f_k}{\partial t}= I(t)_{k,h,f_h (t)} \ ,
$
 where $I$ is the matrix that represents the flux generated by a
move~\cite{Wei02}.

We now separate $A$ in the blocks $S$ and $T$ of constrained and free
variables respectively, writing $A= [S|T]$. The variables that appear in $T$
have an outgoing edge but no incoming ones.  $S$ has $\gamma\: N$ columns,
while $T$ has $(1-\gamma)\:N$ columns.  All the rows of $A$ have $p$ ones. The
distribution for the ones appearing in the columns, i.e. for the outdegree
$k$, is Poisson for both $S$ and $T$, $f_k(0) =
\frac{\lambda^k}{k!}e^{-\lambda}$, with $\lambda(0)= p \gamma$.  We impose
$s_{i,i}=1$.  The lines of $A$ contain on average $p \gamma$ elements in $S$
and $(1-\gamma)p$ elements in $T$, thus after one move there are on average $p
\gamma+ 1$ elements in $S$.  Defining $p'= p\:\gamma+1$. The flux equations
can be written as
\begin{displaymath}
  \begin{array}{l}
    \frac{df^S_k}{dt} = \frac{p'-1}{<k>^S(t)-1}[k f^S_{k+1}- (k-1)
    f^S_k] ;  \quad \mathrm{for~} k>1 \vspace{0.1cm} \\
    \frac{df^S_1}{dt} = -1 + \frac{p'-1}{<k>^S(t)-1}[f^S_{2}] \ \ , \\
\vspace{0.1cm}
    \frac{df^S_0}{dt} = 1 \ \ ,
  \end{array}
\end{displaymath}

where $<k>(t)= \sum k f_k(t)$.  Summing the above equations, one
obtains the evolution equation for the normalization factor $m^S :=
\sum p^c_k =<k>^S-1$. 
$$ 
\frac{d m^S}{dt}= - \frac{p'-1}{m^S(t)-1}\sum (k-1)f^S_k =-p \gamma
$$
With initial condition $m^S(0)=\lambda(0)= p \gamma$, the solution is
$m^S(t)= p \gamma\: (1-t)$. $m^S(t)$ can then be identified with $\lambda(t)$
appearing in the (Poisson) distribution $f_k(t)$.  $-
\frac{\frac{d\lambda(t)}{dt}}{\lambda(t)}= \frac{p'-1}{m^S(t)}=\frac{p
  \gamma}{p \gamma}\: \frac{1}{1-t}$, from which
$\frac{\lambda(t)}{\lambda(0)}= [1-t]$.
Thus, for  $ k>1$,
$$f^S_k= e^{\lambda(t)}\: \frac{\lambda(t)^{k-1}}{(k-1)!}\ \ .$$
For $ k=1$, one can then write $\frac{\partial}{\partial t}f^S_1= -1 -
\frac{\frac{d\lambda}{dt}}{\lambda}\:(\lambda e^{-\lambda})$, so
that 
$$f_1^S(t)= -t + e^{\lambda(t)}= -t+ e^{p \gamma(t-1)}\ \ .$$
The stop time
$t^*$ of the algorithm is then a solution of the equation $t^{*} = e^{p
  \gamma(t^{*}-1)}$.  This last equation implies that if $p \gamma <1$ the
lowest solution for the stop-time is $ t^*=1$, or, in other words, all the
graph is removed. On the other hand, when $ p \gamma >1$, there is a finite
stop time $ t^*<1$, and thus a core. This determines the critical value
$\gamma_{d}^{a} = 1/p$.  The size of the portion of the core matrix contained
in $S$ is given by $M^S_{stop} = N^S_{stop} = \gamma N \:(1-t^*)$.

In order to evaluate the full core matrix and the order parameters, the same
analysis has to be carried out for the matrix of the free variables, $T$. In
this case, one has $p_k^T= k \frac{f^T_k}{m^T(t)}$, where $m^T(t)= \sum k
f_k^T$. Again,
$$\Delta N^T = N(1-\gamma) \frac{\partial}{\partial t} f_k^T \Delta t=
\frac{1-\gamma}{\gamma}\frac{\partial}{\partial t} f_k^T   \ ,      $$
and the flux equations are
\begin{displaymath}
  \begin{array}{l}
\frac{1-\gamma}{\gamma}\frac{\partial}{\partial t} f_0^T =
\frac{p(1-\gamma)}{m^T(t)} f_1^T \ \ , \vspace{0.1cm}\\
\frac{1-\gamma}{\gamma}\frac{\partial}{\partial t} f_1^T =
\frac{p(1-\gamma)}{m^T(t)} [2 f_2^T-f_1^T] \ \ , \vspace{0.1cm} \\
\frac{1-\gamma}{\gamma}\frac{\partial}{\partial t} f_k^T =
\frac{p(1-\gamma)}{m^T(t)} [(k+1) f_{k+1}^T-k f_k^T] . 
  \end{array}
\end{displaymath}
The last equation can be rewritten as
$\frac{\partial}{\partial t} f_k^T = \frac{p \gamma}{m^T(t)} [(k+1)
f_{k+1}^T- k f_k^T]$.
As above, summation yields the evolution of the
normalization constant $\frac{\partial}{\partial t} m^T(t) = -p \gamma$.

\noindent
Thus, $\frac{\frac{\partial}{\partial t} \lambda(t)^T}{\lambda^T}= \frac{p
  \gamma}{m^T(0)-p\gamma t}$, which gives
$$ \frac{ \lambda(t)^T}{\lambda^T(0)}= \frac{m^T(0)-p\gamma t}{m^T(0)}\ \ ,$$
$$f^T_0(\lambda)= e^{-\lambda}\ \ .$$

In conclusion, the stop time $t^*$ is a function of $(p \gamma)$, determined
by the relation $ t^* = e^{p \gamma(t^*-1)}$.  The transition value to an
extensive core is then given by $\gamma_{d}^{a} = 1/ p$.  The core dimensions
can be written as $M^S_C=N\: \gamma (1-t^*)$, and $N^T_C=
(1-\gamma)(1-f_0^T)=(1-\gamma)(1-t^*)$. This last quantity gives the core
order parameter $ \Delta_C = (1-\gamma)(1-t^*)$. $\Delta_{C}$ is zero for
$\gamma < \gamma_{d}^{a}$, and becomes nonzero at this critical value, in a
continuous, non-differentiable way (with an infinite jump). 
The other threshold is easily calculated, as, for any finite $p\gamma$, $t^{*}
> 0$, thus $\gamma_{c}$ is given by the prefactor $1-\gamma$ in $\Delta_{C}$
crossing zero and becoming negative: $\gamma_{c} = 1$.  


\end{appendix}

\end{document}